\documentclass{aip-cp}
\usepackage[T1]{fontenc}
\usepackage[latin9]{inputenc}

\usepackage[numbers]{natbib}
\usepackage{rotating}
\usepackage{graphicx}

\usepackage{savesym}
\savesymbol{iint}
\savesymbol{iiint}
\savesymbol{iiiint}
\savesymbol{idotsint}

\usepackage{amsmath}

\restoresymbol{AMS}{iint}
\restoresymbol{AMS}{iiint}
\restoresymbol{AMS}{iiiint}
\restoresymbol{AMS}{idotsint}

\usepackage{stmaryrd}
\begin{document}

\title{Stochastic Wave-Function Simulation of Irreversible Emission Processes
for Open Quantum Systems in a Non-Markovian Environment}

\author[aff1,aff2]{Evgeny A. Polyakov\corref{cor1}}
\author[aff1,aff3]{Alexey N. Rubtsov}

\affil[aff1]{Russian Quantum Center, 100 Nonaya St., Skolkovo, Moscow 143025,
Russia}

\affil[aff2]{Faculty of Physics, Saint Petersburg State University, 7/9
Universitetskaya Naberezhnaya, Saint Petersburg 199034, Russia}

\affil[aff3]{Department of Physics, Lomonosov Moskow State University,
Leninskie gory 1, 119991 Moscow, Russia}

\corresp[cor1]{Corresponding author: evgenii.poliakoff@gmail.com}

\maketitle

\begin{abstract}
When conducting the numerical simulation of quantum transport, the
main obstacle is a rapid growth of the dimension of entangled Hilbert
subspace. The Quantum Monte Carlo simulation techniques, while being
capable of treating the problems of high dimension, are hindered by
the so-called \textquotedblleft sign problem\textquotedblright . In
the quantum transport, we have fundamental asymmetry between the processes
of emission and absorption of environment excitations: the emitted
excitations are rapidly and irreversibly scattered away. Whereas only
a small part of these excitations is absorbed back by the open subsystem,
thus exercising the non-Markovian self-action of the subsystem onto
itself. We were able to devise a method for the exact simulation of
the dominant quantum emission processes, while taking into account
the small backaction effects in an approximate self-consistent way.
Such an approach allows us to efficiently conduct simulations of real-time
dynamics of small quantum subsystems immersed in non-Markovian bath
for large times, reaching the quasistationary regime. As an example
we calculate the spatial quench dynamics of Kondo cloud for a bozonized
Kodno impurity model.
\end{abstract}

\section{INTRODUCTION}

Real-time dynamics of open quantum systems is actively studied in
such diverse fields as quantum information/computing, solid state
nanodevices, quantum biology and chemical physics \cite{Vega2017}.
The specific type of open quantum system, the Anderson impurity model,
is also highly important in condensed matter physics, since it is
a key component of the dynamical mean-field theory calculations \cite{Freericks2006,Aoki2014}.
This stimulates the development of novel techniques for simulation
of real-time open quantum dynamics. Currently these developments have
resulted in the appearance of efficient simulation methods in such
cases when: the total amount of excitations in the environment is
not large; the memory of the environment is short; the system-environment
coupling is weak; or the spectral density of the environment has special
(Lorentzian) shape \cite{Vega2017}. Nevertheless, recent experimental
advances in fabrication of dispersion-engineered metamaterials urges
us to develop techniques which are capable of handling the cases when:
the memory of environment is long; the system-environment coupling
is strong; we are interested in a long-time dynamics of a driven open
system (or environment at finite temperature), so that the total amount
of emitted/scattered excitations in the environment is unbounded. 

There is no difficulty in simulating the relaxation dynamics of an
excited open quantum system into the vacuum environment. In this case,
only a few excitation quanta are emitted in the environment. Therefore,
we may restrict the total Hilbert space of the environment to the
several-quanta subspace, and then solve the Schrodinger equation exactly
\cite{Vega2017}. However, when there is driving of the open system,
or when the environment is at finite temperature/chemical potential,
the simulation quickly becomes intractable by exact numerical methods.
The reason is that in the first case (driving) the number of emitted
excitations grows (asymptotically) linearly with time, whereas in
the second case (finite temperature/chemical potential) the number
of scattered excitations is linearly growing. Since the dimension
of the relevant Hilbert subspace grows combinatorially with the number
of excitations, the dimension of the problem grows exponentially with
time, and thus we quickly run out of available memory and time resources.

In this work we investigate the idea that the growth of emitted/scattered
environment excitations is the major factor of complexity of real-time
quantum dynamics simulations. A way to eliminate this factor is considered.
The environment excitations are divided into the two kinds: those
which are irreversibly emitted/scattered (``output'' quantum field
of the environment) and those which will be eventually absorbed back
by the open subsystem (quantum field of virtual excitations of the
environment). We propose to conduct a numerically exact stochastic
simulation of the output quantum field, and to evaluate quantum mechanically
the virtual part withn a certain approximation.

There is development in the literature which is similar in spirit:
hybrid stochastic hierarchy equations of motion approach (sHEOM) \cite{Moix2013}.
However, the latter is limited to the Drude-Lorentz spectrum of environment,
whereas our approach in principle may be applied to any spectral shape.
Another distingction of our approach is that we make a physically-informed
separation of the full quantum problem into the exact stochastic and
the approximate deterministic parts.

In section \ref{sec:QUANTUM-FIELD-OF} we provide a precise meaning
for the definitions of the irreversibly emitted output field and of
the virutal excitation field. It is shown that the interaction with
the output quantum field can be efficiently simulated without the
sign problem by a stochastic wavefunction method analogous to that
of Diosi, Gisin, and Strunz \cite{Diosi1997,Diosi1998}. However,
the important disctinction from the latter is that we explicitly keep
the full quantum mechanical problem for virtual excitations. In section
\ref{sec:ZERO-ORDER-APPROXIMATION-TO} we discuss the approximations
to the quantum virtual field and introduce the simplest zero-order
approximation. Then, in section \ref{sec:BOSONIZATION-OF-KONDO} we
apply our approach to the bozonized Kondo model and compute the spatial
quench dynamics of the Kondo cloud. Finally, a conclusion is made
in section \ref{sec:CONCLUSION}.

\section{\label{sec:QUANTUM-FIELD-OF}QUANTUM FIELD OF ENVIRONMENT: CLASSICAL
INTERPRETATION OF THE IRREVERSIBLY EMITTED AND THERMAL PARTS}

We conder the following Hamiltonian
\begin{equation}
\widehat{H}=\widehat{H}_{\textrm{sys}}+\hat{a}\widehat{b}^{\dagger}+\hat{a}^{\dagger}\widehat{b}+\widehat{H}_{\textrm{env}},
\end{equation}
where $\widehat{H}_{\textrm{env}}$ is an enviroment with quadratic
Hamiltonian,
\begin{equation}
\widehat{H}_{\textrm{env}}=\intop_{-\infty}^{+\infty}d\omega\omega\widehat{b}^{\dagger}\left(\omega\right)\widehat{b}\left(\omega\right),
\end{equation}
which is coupled bilinearly to the open system $\widehat{H}_{\textrm{sys}}$
through a certain system operator $\widehat{a}$ and the environment's
collective degree of freedom $\widehat{b}$, 
\begin{equation}
\widehat{b}=\intop_{-\infty}^{+\infty}d\omega c\left(\omega\right)\widehat{b}\left(\omega\right).
\end{equation}
Observe that in our representation the frequency dependence of the
density-of-states is transfered to the coefficient $c\left(\omega\right)$. 

Suppose that initially the open system is in a pure state $\left|\psi_{\textrm{sys}}\right\rangle $,
and the environment is initially in a termal state at inverse temperature
$\beta$, with the mode occupations
\begin{equation}
n\left(\omega\right)=\textrm{Tr}\left\{ \widehat{b}^{\dagger}\left(\omega\right)\widehat{b}\left(\omega\right)\widehat{\rho}_{\textrm{env}}\right\} .
\end{equation}
 For the real-time evolution during time interval $\left[0,t\right]$,
we may write the Keldysh contour partition function as a formal path
integral over the forward/backward fields of open system $a_{\pm}\left(\tau\right)$,
and over the environment $b_{\pm}\left(\tau\right)$,
\begin{equation}
Z\left[0,t\right]=\int D\left[a_{\pm}\right]D\left[b_{\pm}\right]\exp\left(S\left[a_{\pm},b_{\pm}\right]\right),
\end{equation}
with an appropriate Keldysh action $S\left[a_{\pm},b_{\pm}\right]$
\cite{Kamenev2011}. We integrate out the environment's degrees of
freedom $b_{\pm}\left(\tau\right)$, and obtain
\begin{equation}
Z\left[0,t\right]=\int D\left[a_{\pm}\right]\exp\left(S_{\textrm{eff}}\left[a_{\pm}\right]\right),
\end{equation}
where the resulting effective action $S_{\textrm{eff}}\left[a_{\pm}\right]$
has the form 
\begin{equation}
S_{\textrm{eff}}\left[a_{\pm}\right]=S_{\textrm{sys}}\left[a_{\pm}\right]-\iintop_{0}^{T}d\tau d\tau^{\prime}\left[\begin{array}{c}
-a_{+}^{*}\left(\tau\right)\\
a_{-}^{*}\left(\tau\right)
\end{array}\right]^{T}\mathbf{K}\left(\tau-\tau^{\prime}\right)\left[\begin{array}{c}
-a_{+}\left(\tau^{\prime}\right)\\
a_{-}\left(\tau^{\prime}\right)
\end{array}\right],
\end{equation}
and in the second hybridization term the matrix $\mathbf{K}\left(\tau-\tau^{\prime}\right)$
is the Keldysh function of the bath. We divide it into three parts:
\begin{equation}
\mathbf{K}\left(t-t^{\prime}\right)=\mathbf{K}_{\textrm{virt}}\left(t-t^{\prime}\right)+\mathbf{K}_{\textrm{emit}}\left(t-t^{\prime}\right)+\mathbf{K}_{\textrm{therm}}\left(t-t^{\prime}\right),
\end{equation}
where the term
\begin{equation}
\mathbf{K}_{\textrm{virt}}\left(t-t^{\prime}\right)=\left[\begin{array}{cc}
\theta\left(t-t^{\prime}\right) & 0\\
0 & \theta\left(t^{\prime}-t\right)
\end{array}\right]M\left(t-t^{\prime}\right)
\end{equation}
describes the effect of the virtual environmental excitations (like
Lamb shift);
\begin{equation}
\mathbf{K}_{\textrm{emit}}\left(t-t^{\prime}\right)=\left[\begin{array}{cc}
0 & 0\\
1 & 0
\end{array}\right]M\left(t-t^{\prime}\right)
\end{equation}
describes the irreversible emission of observable excitations;
\begin{equation}
\mathbf{K}_{\textrm{therm}}\left(t-t^{\prime}\right)=\left[\begin{array}{cc}
1 & 1\\
1 & 1
\end{array}\right]M_{\textrm{therm}}\left(t-t^{\prime}\right)
\end{equation}
represents the effects of themal fluctuations of environment. Here
the environment memory function is
\begin{equation}
M\left(t-t^{\prime}\right)=\intop_{-\infty}^{+\infty}d\omega\left|c\left(\omega\right)\right|^{2}e^{-i\omega\left(t-t^{\prime}\right)},
\end{equation}
and the termal noise memory function is
\begin{equation}
M_{\textrm{therm}}\left(t-t^{\prime}\right)=\intop_{-\infty}^{+\infty}d\omega n\left(\omega\right)\left|c\left(\omega\right)\right|^{2}e^{-i\omega\left(t-t^{\prime}\right)}.
\end{equation}
In Fig. \ref{fig:Fig_Keldysh_perturbative} we present the hybridization
expansion of partition function with respect to $\mathbf{K}_{\textrm{virt}}\left(t-t^{\prime}\right)$
and $\mathbf{K}_{\textrm{emit}}\left(t-t^{\prime}\right)$, which
is the basis for our physical interpretation. It is seen that $\mathbf{K}_{\textrm{virt}}\left(t-t^{\prime}\right)$
corresponds to the contractions inside one branch of Keldysh contour
(virtual excitations which are ultimately absorbed back), and $\mathbf{K}_{\textrm{emit}}\left(t-t^{\prime}\right)$
corresponds to contractions across different branches (irreversibly
emitted excitations which contribute to the trace over the environment's
states). It turns out that $\mathbf{K}_{\textrm{emit}}\left(t-t^{\prime}\right)$
and $\mathbf{K}_{\textrm{therm}}\left(t-t^{\prime}\right)$can be
interpreted in a classical way. Indeed, in the spirit of stochastic
wavefunction method of Diosi, Gisin, and Strunz \cite{Diosi1997,Diosi1998},
we introduce a $c$-number stochastic fields
\begin{equation}
z_{\textrm{emit}}\left(t\right)=\intop_{-\infty}^{+\infty}d\omega c\left(\omega\right)e^{-i\omega t}\xi\left(\omega\right)
\end{equation}
and
\begin{equation}
v_{\textrm{therm}}\left(t\right)=\intop_{-\infty}^{+\infty}d\omega\sqrt{n\left(\omega\right)}c\left(\omega\right)e^{-i\omega t}\eta\left(\omega\right),
\end{equation}
where $\xi\left(\omega\right)$ and $\eta\left(\omega\right)$ are
complex independent white noises:
\begin{equation}
\overline{\xi^{*}\left(\omega\right)\xi\left(\omega^{\prime}\right)}=\delta\left(\omega-\omega^{\prime}\right),\,\,\,\overline{\eta^{*}\left(\omega\right)\eta\left(\omega^{\prime}\right)}=\delta\left(\omega-\omega^{\prime}\right).
\end{equation}
The following stochastic Hamiltonian is introduced
\begin{equation}
\widehat{H}_{\textrm{stoch}}\left(z_{\textrm{emit}}\left(t\right),v_{\textrm{therm}}\left(t\right)\right)=\widehat{H}_{\textrm{sys}}+\hat{a}\left\{ \widehat{b}^{\dagger}+z_{\textrm{emit}}^{*}\left(t\right)+v_{\textrm{therm}}^{*}\left(t\right)\right\} +\hat{a}^{\dagger}\left\{ \widehat{b}+v_{\textrm{therm}}\left(t\right)\right\} +\widehat{H}_{\textrm{env}}.
\end{equation}
Here we see that the effect of the finite temperature can be exactly
represented as a classical fluctuating field $v_{\textrm{therm}}\left(t\right)$.
Then, since the termal bath state is Gaussian, when we calculate the
averages, the Wick theorem rules applies for all the occurences of
the bath operators $\widehat{b}\left(t\right)$ and $\widehat{b}^{\dagger}\left(t\right)$.
However, according to the stochastic wavefunction method, the Wick
theorem rules do not change if we replace $\widehat{H}$ by $\widehat{H}_{\textrm{stoch}}\left(t\right)$
and instead of the full trace we compute the environment vacuum-vacuum
amplitude: 
\begin{equation}
Z\left[0,t\right]=\overline{\left|\left\langle 0_{\textrm{env}},\psi_{\textrm{sys}}\right|\mathcal{T}\exp\left[-i\intop_{0}^{t}d\tau\widehat{H}_{\textrm{stoch}}\left(z_{\textrm{emit}}\left(\tau\right),v_{\textrm{therm}}\left(\tau\right)\right)\right]\left|0_{\textrm{env}},\psi_{\textrm{sys}}\right\rangle \right|^{2}}_{z_{\textrm{emit}},v_{\textrm{therm}}}.\label{eq:stochastic_partition_function}
\end{equation}
Note that we consider the virtual excitations component $\mathbf{K}_{\textrm{virt}}\left(t-t^{\prime}\right)$
as a genuinely quantum object. Indeed, all the attempts to employ
a stochastic sampling of $\mathbf{K}_{\textrm{virt}}\left(t-t^{\prime}\right)$
lead to an exponential growth of complexity of calculations, numerical
instabilities, etc, which we call a ``generalized sign problem''.
Therefore, we propose to compute the influence of $\mathbf{K}_{\textrm{virt}}\left(t-t^{\prime}\right)$
quantum-mechanically, within a ceratin approximation. 

The exact stochastic evaluation of real-time dynamics of a certain
observable system $\widehat{O}_{\textrm{sys}}$ proceeds in the following
way. We solve the stochastic Schrodinger equation \textit{with bath
degrees of freedom} 
\begin{equation}
\partial_{t}\left|\Psi\left(t\right)\right\rangle =-i\widehat{H}_{\textrm{stoch}}\left(z_{\textrm{emit}}\left(t\right),v_{\textrm{therm}}\left(t\right)\right)\left|\Psi\left(t\right)\right\rangle ,\label{eq:stochastic_Schrodinger_equation}
\end{equation}
for a given realization of noises $z_{\textrm{emit}}$ and $v_{\textrm{therm}}$
and for the initial condition $\left|0_{\textrm{env}},\psi_{\textrm{sys}}\right\rangle $.
Then, the observable average is calculated as 
\begin{equation}
\left\langle \widehat{O}_{\textrm{sys}}\right\rangle =\overline{\left\langle \Psi\left(t\right)\right|\left.0_{\textrm{env}}\right\rangle \widehat{O}_{\textrm{sys}}\left\langle 0_{\textrm{env}}\right.\left|\Psi\left(t\right)\right\rangle }_{z_{\textrm{emit}}\left(t\right),v_{\textrm{therm}}\left(t\right)}.\label{eq:system_observable}
\end{equation}
From the commutation relations it can also be shown that the environment's
observables $\widehat{O}_{\textrm{env}}$ in the antinormally ordered
form can be computed by subsituting 
\begin{equation}
\widehat{b}\left(\omega\right)\to e^{-i\omega t}\left(\xi\left(\omega\right)+\sqrt{n\left(\omega\right)}\eta\left(\omega\right)\right),
\end{equation}
\begin{equation}
\widehat{b}^{\dagger}\left(\omega\right)\to e^{i\omega t}\left(\xi^{*}\left(\omega\right)+\sqrt{n\left(\omega\right)}\eta^{*}\left(\omega\right)\right)
\end{equation}
in the operator expression for $\vdots\widehat{O}_{\textrm{env}}\vdots$,
obtaining a $c$-number function $O_{\textrm{env}}\left(\xi,\eta\right)$.
Then, the average is computed as 
\begin{equation}
\left\langle \vdots\widehat{O}_{\textrm{env}}\vdots\right\rangle =\overline{\left\langle \Psi\left(t\right)\right|\left.0_{\textrm{env}}\right\rangle O_{\textrm{env}}\left(\xi,\eta\right)\left\langle 0_{\textrm{env}}\right.\left|\Psi\left(t\right)\right\rangle }_{z_{\textrm{emit}}\left(t\right),v_{\textrm{therm}}\left(t\right)}.
\end{equation}
Note that this ordering prescription closely resembles the Husimi
Q-function representation of the environemnt's state (in the interaction
picture).

\section{\label{sec:ZERO-ORDER-APPROXIMATION-TO}ZERO-ORDER APPROXIMATION
TO VIRTUAL EXCITATIONS}

\begin{figure}
\includegraphics[scale=0.6]{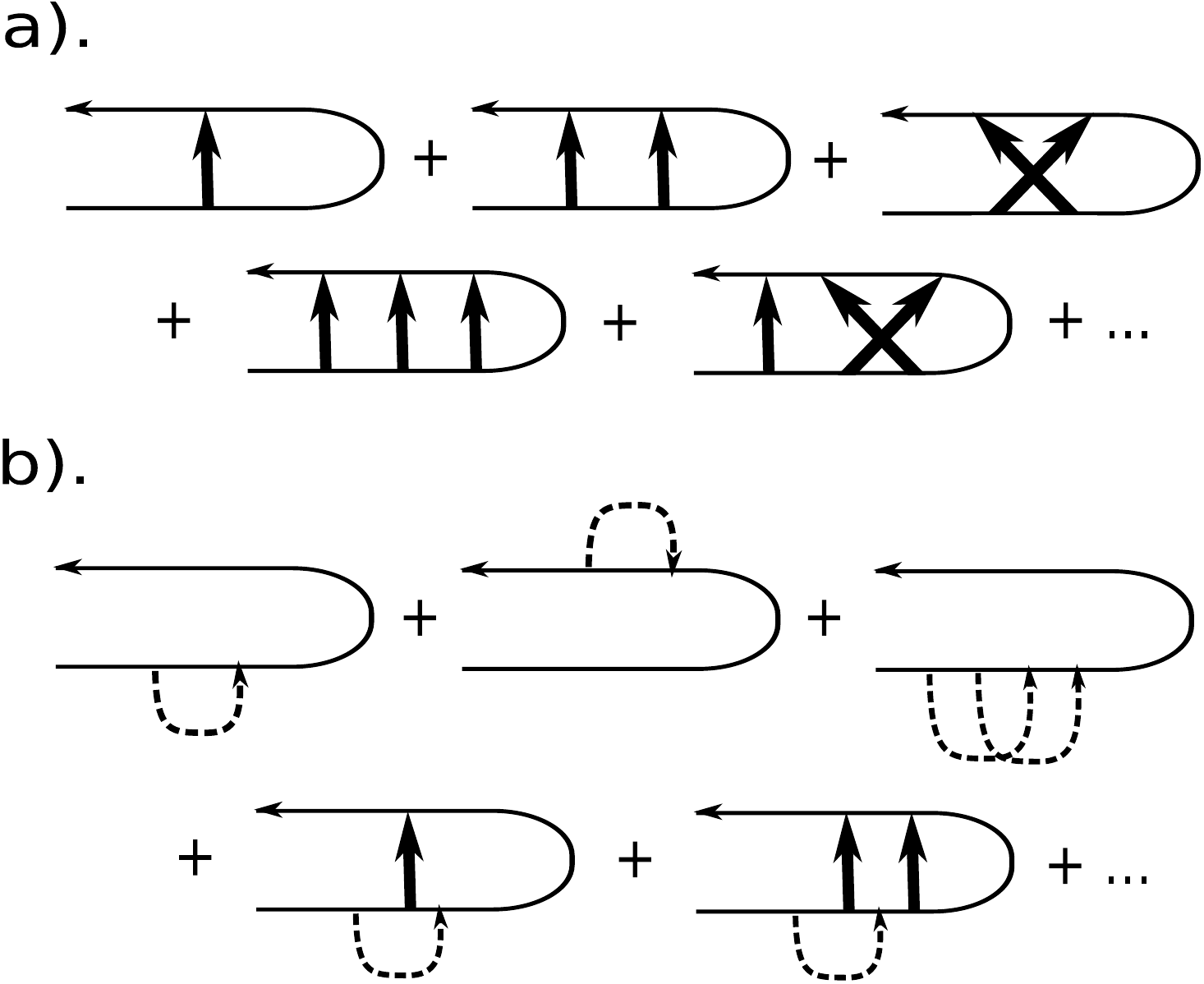}

\caption{\label{fig:Fig_Keldysh_perturbative}Two types of diagrams in the
perturbative expansion of open system evolution on the Keldysh contour.
a). Cross-branch diagrams corresponding to the irreversible emission
of observable environment excitation. b). The diagrams containing
intra-branch contractions, corresponding to the emission and absorption
of unobservable virtual excitations. }

\end{figure}

The simplest approximation is to completely neglect the intrabranch
contractions $\mathbf{K}_{\textrm{virt}}\left(t-t^{\prime}\right)$.
This corresponds to neglecting all the diagrams with the intrabranch
contractions in Fig. \ref{fig:Fig_Keldysh_perturbative}, and neglecting
completely the environment degrees of freedom in the stochastic Schrodinger
equation (\ref{eq:stochastic_Schrodinger_equation}). We obtain the
following stochastic recipe:
\begin{equation}
\partial_{t}\left|\psi_{\textrm{sys}}\left(t\right)\right\rangle =-i\widehat{H}_{\textrm{sys}}\left|\psi_{\textrm{sys}}\left(t\right)\right\rangle -i\hat{a}\left\{ z_{\textrm{emit}}^{*}\left(t\right)+v_{\textrm{therm}}^{*}\left(t\right)\right\} \left|\psi_{\textrm{sys}}\left(t\right)\right\rangle -i\hat{a}^{\dagger}v_{\textrm{therm}}\left(t\right)\left|\psi_{\textrm{sys}}\left(t\right)\right\rangle .\label{eq:zero_order_stochastic_trajectories}
\end{equation}
Note that the virtual excitations play the following two roles. a).
by drawing the analogy with Lindblad equation in the Markovian case,
we argue that they ensure the balance of probability density distribution
of quantum trajectories: they vary the norm of the quantum trajectory
to take into account the probability flux into other trajectories.
b). they take into account the dynamical effects of virtual excitations,
e.g. Lamb shift of levels. Therefore, by discarding these virtual
terms, we break a). and b). In order to fix approximately a)., when
computing observables for the trajectories Eq. (\ref{eq:zero_order_stochastic_trajectories}),
we divide by the average norm of the enseble of trajectories:
\begin{equation}
\left\langle \widehat{O}_{\textrm{sys}}\right\rangle =\frac{\overline{\left\langle \psi_{\textrm{sys}}\left(t\right)\right.\widehat{O}_{\textrm{sys}}\left.\psi_{\textrm{sys}}\left(t\right)\right\rangle }_{z_{\textrm{emit}},\nu_{\textrm{therm}}}}{\overline{\left\langle \left.\psi_{\textrm{sys}}\left(t\right)\right|\psi_{\textrm{sys}}\left(t\right)\right\rangle }_{\textrm{z}_{\textrm{emit}},\nu_{\textrm{therm}}}}.\label{eq:system_observable-1}
\end{equation}
These two equations, (\ref{eq:zero_order_stochastic_trajectories})
and (\ref{eq:system_observable-1}), is the resulting zero-order approximation
(without virtual excitations).

\section{\label{sec:BOSONIZATION-OF-KONDO}BOSONIZATION OF KONDO MODEL}

\begin{figure}
\includegraphics[scale=0.52]{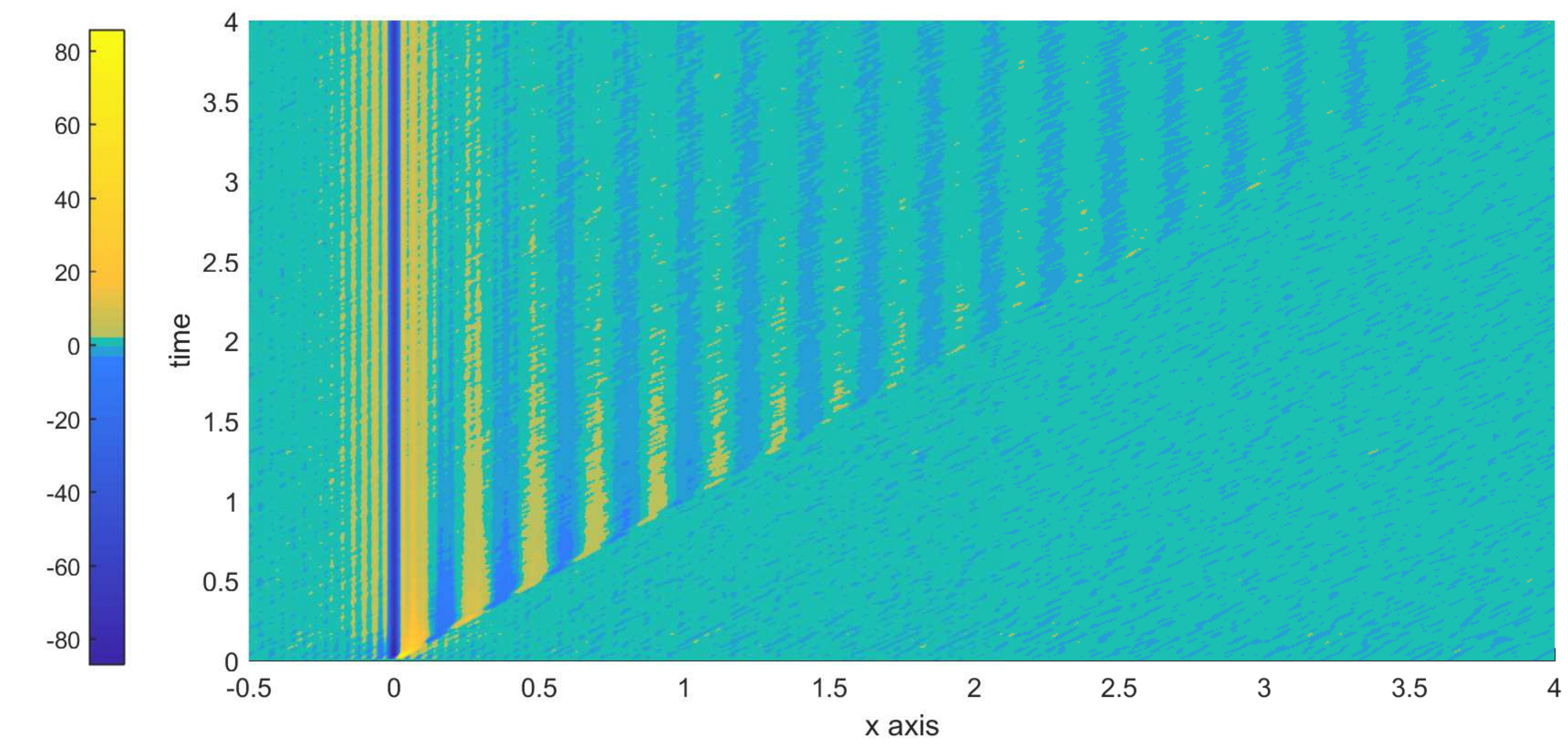}

\caption{\label{fig:quench_of_spin_boson_model}The time and spatially dependent
spin-spin correlation function $\left\langle \sigma_{z}\left(t\right)S_{z}\left(x,t\right)\right\rangle $
between the impurity and the Kondo cloud as a color contour plot.
After the initial coupling is switched on at $t=0$, the correlation
propagates with the Fermi velocity $v_{F}$. }

\end{figure}

In this section we apply our method to the spin-1/2 Kondo model which
describes a localized magnetic moment interacting with electron gas
\cite{Nghiem2016,Guinea1985,Schlottmann1982,Delft1998}. The characteristic
property of this model is that in the ground state the electron gas
becomes correlated with the impurity, the so called Kondo screening
cloud, which forms a spin singlet round impurity. The 1d chiral Kondo
problem has the following Hamiltonian
\begin{equation}
\widehat{H}=\sum_{k\mu}\varepsilon_{k}c_{k\mu}^{\dagger}c_{k\mu}+\frac{J_{\perp}}{2}\sum_{k,k^{\prime}}\left(c_{k\shortuparrow}^{\dagger}c_{k^{\prime}\shortdownarrow}S_{\textrm{imp}}^{-}+c_{k\shortdownarrow}^{\dagger}c_{k^{\prime}\shortuparrow}S_{\textrm{imp}}^{+}\right)+\frac{J_{\bigparallel}}{2}\left(c_{k\shortuparrow}^{\dagger}c_{k^{\prime}\shortuparrow}-c_{k\shortdownarrow}^{\dagger}c_{k^{\prime}\shortdownarrow}\right)S_{\textrm{imp}}^{z},
\end{equation}
where 
\begin{equation}
k=2\pi n/L,\,\,\,-\infty<n<\infty,
\end{equation}
$L$ is the quantization volume, $L\to\infty$. The 1d chiral Kondo
is mapped onto the spin-boson model \cite{Nghiem2016,Guinea1985,Schlottmann1982,Delft1998}
\begin{equation}
\widehat{H}=\frac{J_{\perp}}{4\pi a}\sigma_{x}+v_{F}\sum_{q}qb_{q}^{\dagger}b_{q}+\left(\frac{J_{\bigparallel}}{4\pi}-v_{F}\right)\sqrt{2}\frac{\sigma_{z}}{2}\sum_{q>0}\sqrt{\frac{2\pi q}{L}}\left(b_{q}+b_{q}^{\dagger}\right)e^{-aq/2},\label{eq:spin_boson}
\end{equation}
for the spectral regularization parameter $a\to0$. 
\begin{equation}
q=2\pi n_{q}/L,\,\,\,0<n_{q}<\infty,
\end{equation}
The bosonic operators are related to the fermionic ones as
\begin{equation}
b_{q\mu}^{\dagger}=in_{q}^{-1/2}\sum_{k}c_{k+q\mu}^{\dagger}c_{k\mu}.
\end{equation}
In this work we employed the following parameters: $\left|n\right|\leq1600$,
$L=60$, $J_{\perp}=J_{\bigparallel}=2\pi\times0.3$, $v_{F}=1,\,\,\,a=0.01$.
According to the approximate stochastic algorithm (\ref{eq:zero_order_stochastic_trajectories})-(\ref{eq:system_observable-1}),
we have the open system and the environment Hamiltonians
\begin{equation}
\widehat{H}_{\textrm{sys}}=\frac{J_{\perp}}{4\pi a}\sigma_{x},\,\,\,\widehat{H}_{\textrm{env}}=v_{F}\sum_{q}qb_{q}^{\dagger}b_{q}.
\end{equation}
The system is coupled through its operator
\begin{equation}
\widehat{a}=\sigma_{z},
\end{equation}
and the bath's collective degree of freedom 
\begin{equation}
\widehat{x}=\left(\frac{J_{\bigparallel}}{4\pi}-v_{F}\right)\sum_{q>0}\sqrt{\frac{2\pi q}{L}}\left(b_{q}+b_{q}^{\dagger}\right)e^{-aq/2}.
\end{equation}
Letus take the vacuum initial state of the environmnet, and the impurity
spin initially directed towards $z$-axis. We switch on the coupling
at $t=0$ and conduct the stochastic Monte Carlo simulation of the
resulting quench dynamics. In Fig. \ref{fig:quench_of_spin_boson_model}
we present the results of calculation of spin-spin correlation function
$\left\langle \sigma_{z}\left(t\right)S_{z}\left(x,t\right)\right\rangle $.

\begin{figure}
\includegraphics[scale=0.52]{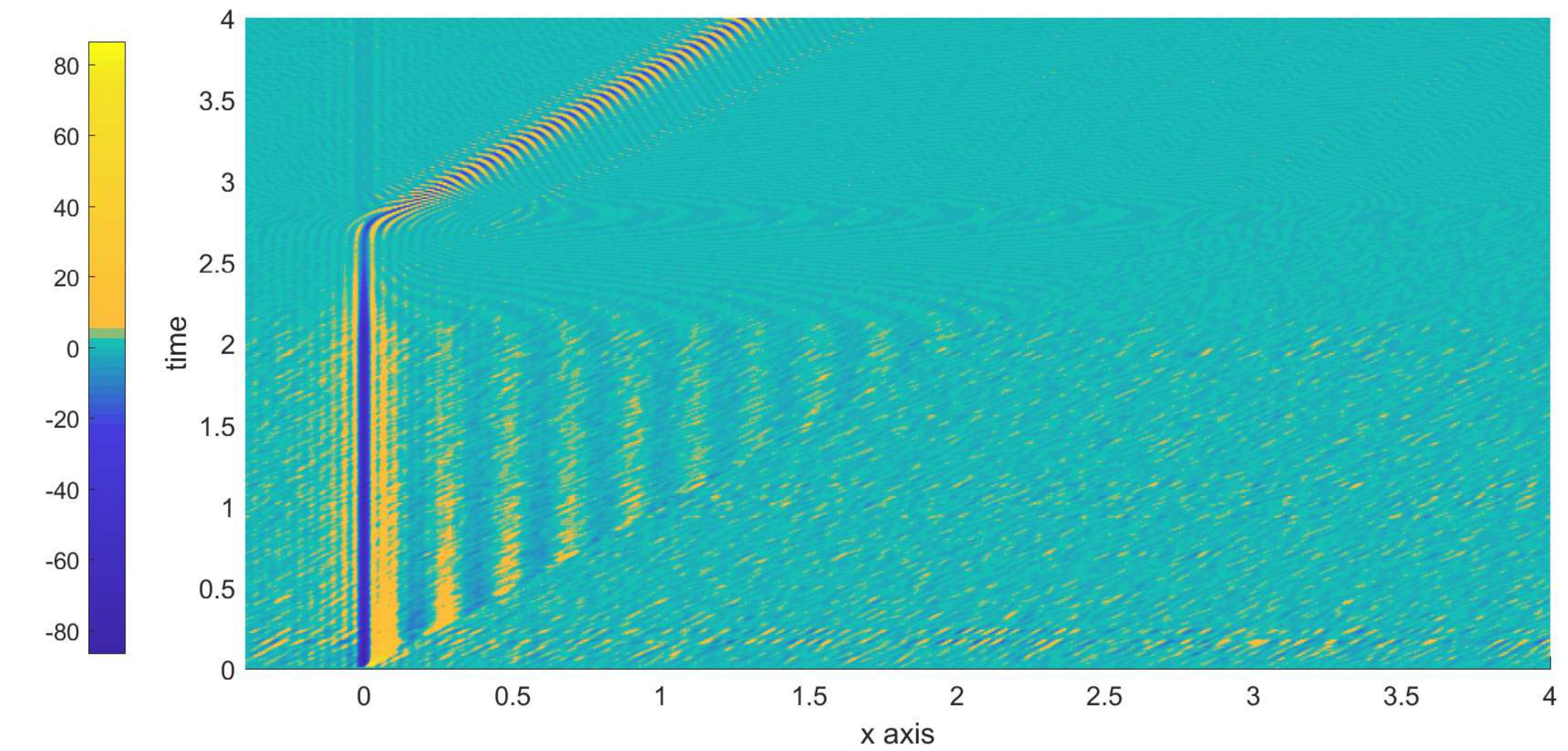}

\caption{\label{fig:kondo_cloud_flies_away}The time and spatially dependent
spin-spin correlation function $\left\langle \sigma_{z}\left(t\right)S_{z}\left(x,t\right)\right\rangle $
between the impurity and the Kondo cloud as a color contour plot.When
the orientational fluctuations of the impurity spin are supressed
by a strong external field at time moment around $t=3$, the Kondo
cloud cannot be sustained, it detaches from the impurity and flies
away at the speed $v_{F}$. }

\end{figure}

In Fig. \ref{fig:kondo_cloud_flies_away} we present simulation result
when an additional quench is done: an external field 
\begin{equation}
\widehat{H}_{\textrm{ext}}=500\sigma_{x}
\end{equation}
is smoothly switched on by the time moment $t=3$. Such a strong field
supresses orientational fluctuations of the impurity spin, which are
important for sustaining the Kondo cloud. As a consequence, the Kondo
cloud detaches from the impurity and flies away.

\section{\label{sec:CONCLUSION}CONCLUSION}

In this work we propose the idea of how to advance in the development
of real-time simulation techniques for open quantum systems. Our approach
is to calculate in a numerically exact way all the observable effects
of the environment: irreversibly emitted excitations and the thermal
field. Due to this observability, we obtain true probabilities for
Monte Carlo simulation at long times without the sign problem. However,
we argue that the genuinely quantum virtual excitations cannot be
interpreted in a classical way without the sign problem, and should
be computed quantum mechanically in a certain approximate way. In
this work we demonstrate the simplest approximation: to completely
neglect the virtual terms. On the example of chiral Kondo model we
easily obtain interesting semiquantitative results. In particular,
the stochastic simulation demonstrates how the process of Kondo cloud
formation propagates in space at the Fermi velocity. Another simulated
effect is that if the impurity spin fluctuations are supressed by
a strong external field, the Kondo cloud detaches and flies away.

Higher order approximations for the quantum field of virtual excitations
may be implemented by e. g. including a truncated basis of environment
states into the stochastic Schrodinger equation (\ref{eq:zero_order_stochastic_trajectories});
by employing (multi)polaronic expansion \cite{Bera2014,Bera2014a}
etc.
\section{ACKNOWLEDGMENTS}
The study was founded by the RSF, grant 16-42-01057.

\nocite{*}
\bibliographystyle{aipnum-cp}%

\end{document}